\documentclass[showpacs,twocolumn]{revtex4}

\usepackage{graphicx}
\usepackage{dcolumn}
\usepackage{amsmath}
\usepackage{color}

\makeatletter
\def\btt#1{\texttt{\@backslashchar#1}}
\DeclareRobustCommand\bblash{\btt{\@backslashchar}} \makeatother

\begin{document}

\title[Shorttitle]{Shadow of five-dimensional rotating Myers-Perry black hole}
\author{Uma Papnoi$^{a}$} \email{uma.papnoi@gmail.com}
\author{Farruh Atamurotov$^{b}$} \email{farruh@astrin.uz}
\author{Sushant~G.~Ghosh$^{a,\;c\;}$} \email{sghosh2@jmi.ac.in,
sgghosh@gmail.com}
\author{Bobomurat Ahmedov$^{b,\; d,\; e}$} \email{ahmedov@astrin.uz}
\affiliation{$^{a}$ Centre for Theoretical Physics, Jamia Millia
Islamia, New Delhi 110025, India}
\affiliation{$^{b}$ Institute of Nuclear Physics, Ulughbek, Tashkent 100214, Uzbekistan}%
\affiliation{$^{c}$ Astrophysics and Cosmology
Research Unit, School of Mathematical Sciences, University of
Kwa-Zulu-Natal, Private Bag 54001, Durban 4000, South Africa}
\affiliation{$^{d}$ The Abdus Salam International Centre for Theoretical Physics, 34151 Trieste, Italy}
\affiliation{$^{e}$ Ulughbek Astrononmical Institute, Astronomicheskaya 33, Tashkent 100052, Uzbekistan}
\date{\today}

\begin{abstract}
A black hole casts a shadow as an optical appearance because of
its strong gravitational field. We study the shadow cast by the
five-dimensional Myers-Perry black hole  with equal rotation
parameters.  We demonstrate that the null geodesic equations can
be integrated that allows us to investigate the shadow cast by a
black hole. The shadow of a black hole is found to be a dark zone
covered by deformed circle.   Interestingly, the shapes of the
black hole shadow are more distorted and size decreases for larger
black hole spins. Interestingly, it turns out that, for fixed
values of rotation parameter, the shadow is slightly smaller and
less deformed than for its four-dimensional Kerr black
counterpart. Further, the shadow of the five-dimensional Kerr
black hole is concentric deformed circles.  The effect of rotation
parameter on the shape and size of a naked singularity shadow is
also analyzed.
\end{abstract}

\pacs{04.50.-h, 04.25.-g, 04.70.-s}
\maketitle

\section{Introduction}

It is strongly believed that there exist black holes in the
centres of majority galaxies, for instance, a widely accepted
opinion that radio-source Sgr A* in the galactic centre of Milky
Way is likely to be a supermassive black hole \cite{HM, BR}. Since
the galaxies are rotating, it is very likely that a black hole at
the centre of a galaxy also possesses a spin. There is a great
interest to investigate nature of the black holes, i.e., mass and
spin of the black hole. Observation of black hole  shadow is one
of the possible methods to determine the mass and spin of a
rotating black hole \cite{Vries, Takahashi04, Bambi, Kraniotis}.
Now, it is a general belief that a black hole, if it is in front
of a luminous background, will cast a shadow. The black hole
shadow is the optical appearance casted by a black hole, and its
existence was first studied by Bardeen \cite{JMB}. The study of
black hole shadow received
 a significant attention and has become a quite active research field (for a review, see \cite{bozzareview}).
 The shadow of Schwarzschild black hole \cite{D, otros},
 rotating black hole with gravitomagnetic and electric charge
 \cite{claus1} and other spherically symmetric black holes have been intensively studied.
 The gravitational lensing and optical phenomena are thoroughly investigated
 for the Janis-Newman-Winicour space-time in \cite{boz} and for the Kerr-Newman and and the
Kerr-Newman-(anti) de Sitter
spacetimes in \cite{Kraniotis14}.
 It has been demonstrated that the shadow of the Schwarzschild black hole is a perfect circle \cite{bozzareview} through the gravitational lensing either in vacuum~\cite{vih} or in plasma~\cite{vika}. On the other hand, for rotating case,
 it has elongated shape in the direction of rotation \cite{bozza2, chandra, zakharov05, Ned, Li1, Falcke}.  Shadows of black holes possessing nontrivial NUT charge were obtained in \cite{Abdujabbarov},
 while the Kerr-Taub-NUT black hole was discussed in \cite{Abdujabbarov}
 and black hole solutions within Einstein-Maxwell-dilaton gravity and Chern-Simons gravity was considered in \cite{amar}. The apparent shape of the Sen black hole is studied in \cite{Hioki08},
  and rotating braneworld black holes were investigated in \cite{Amarilla2, schee}. Both Schwarzschild and rotating black hole with an accretion disc were investigated in \cite{jp, stu}.
 Further, the effect of spin parameter and the angle of inclination on
the shape of the shadow was extended to the Kaluza-Klein
rotating dilaton black hole \cite{Amarilla1},  the rotating
Horava-Lifshitz black hole \cite{Fa2}, rotating Non-Kerr black hole \cite{Fa3}
and Einstein-Maxwell-Dilaton-Axion black hole \cite{Wei}.

Black holes are considered as highly interesting gravitational
compact as well as geometrical objects to study in four dimensions
and their existence in higher dimensional space-time is not
excluded. In recent years, black hole solutions in more  than
four space-time dimensions, especially in five-dimensions (5D),
have been the subject of intensive research, motivated by ideas in
brane-world cosmology, string theory and gauge/gravity duality.
Several interesting and surprising results have been found
\cite{horowitz}. In dimensions higher than four, the uniqueness
theorems do not hold due to the fact that there are more degrees
of freedom.  The discovery of black-ring solutions in five
dimensions  shows that non-trivial topologies are allowed in
higher dimensions \cite{Emparan}. The Myers-Perry \cite{MP} black
hole spacetime is higher dimensional generalization of the
four-dimensional Kerr black hole spacetime. In particular,
energetics of a 5D rotating  black hole have been studied
in~\cite{Dadhich10,Tsukamoto13,Abdujabbarov13b}.

The aim of the current paper is to investigate the apparent shape of
a 5D Myers-Perry solution to visualize the shape
of the shadow and compare the results with the images for the Kerr
black hole. Thus, we can draw conclusions on their possible
distinction in astrophysical observation.  The paper is organized as
follows: In Sec. II, we have discussed about the 5D Myers-Perry
black hole solution. In Sec. III, we have presented the particle
motion around the 5D black hole to discuss about the shadow of the
black hole. In Sec. IV and V, two observables are introduced to
discuss about the apparent shape of the shadow of the black hole and
naked singularity and finally in Sec. VI, we have concluded with the results.

The metric considered here is a five-dimensional generalization of
the Kerr metric. We have used units that fix the speed of light and
the gravitational constant via $8\pi G = c^4 = 1$.

\section{Five-Dimensional Myers-Perry black hole}

The metric of a 5D rotating Myers-Perry black hole in the
Boyer-Lindquist coordinates reads \cite{MP}
\begin{eqnarray}\label{Kerr5D}
ds^2 &=& \frac{\rho^2}{ 4\Delta} dx^2+\rho^2
d\theta^2 - dt^2 +(x+a^2) \sin^2\theta d\phi^2 \nonumber \\  && +  (x+b^2)  \cos^2\theta d\psi^2 + \frac{r_0^2}{\rho^2} \Big[dt+a \sin^2\theta d\phi  \nonumber \\  && +b
\cos^2\theta d\psi  \Big]^2, \end{eqnarray}
with $\rho^2$ and $\Delta$ are given by
\begin{eqnarray}
\rho^2=x+a^2\cos^2\theta+b^2 \sin^2\theta , \nonumber \\
\Delta=(x+a^2)(x+b^2)-r_0^2 x. \nonumber
\end{eqnarray}It may be noted that the metric (\ref{Kerr5D}) is singular when $\Delta=g_{rr}=0$ and $\rho^2=0$. Here $a$ and $b$ are two spin parameters, and $0\leq\phi\leq2\pi$ and $0\leq\psi\leq\pi/2$ are two angles. Following \cite{Frolov:2003en} instead of the radius $r$ we use the coordinate $x=r^2$ to simplify the calculations. The ADM mass of black hole is $2M=r_0^2$. Also note that the metric (\ref{Kerr5D}) reduces to 5D Tangherlini solution \cite{scht} for $a=b=0$.

The black hole horizon is determined by taking roots of the equation $\Delta=0$, which admit solutions
\begin{equation}\label{horizon}
x_{\pm}=\frac{1}{2}\left[r_0^2-(a^2+b^2)\pm
\sqrt{[r_0^2-(a^2+b^2)]^2-4a^2b^2}\right] .\nonumber
\end{equation}Here, $x_+$ denotes outer horizon and $x_-$ the inner horizon. It is clear that the metric (\ref{Kerr5D}) describes non-extremal black hole for $x_+>x_-$ and when $x_+=x_-$, one obtains an extremal black hole.  The horizon exists when $a^2+b^2<r_0^2$ and $[r_0^2-(a^2+b^2)]^2\geq4 a^2b^2$. This defines a region in the $(a,b)$ where metric represents black hole and not a naked singularity.

Eq. (\ref{horizon}), when $a=b$, yields
\begin{equation}\label{horizon1}
x_{\pm}=\frac{1}{2}(r_0^2-2 a^2\pm \sqrt{r_0^4-4 a^2 r_0^2}).
\end{equation}
The extremal black hole occurs when $r_0^2=4 a^2$ horizon merges. On the other hand for $r_0^2<4a^2$ we have a naked singularity and black hole occurs when $r_0^2>4 a^2$.  It may be noted that the event horizon $r_{+}$ is smaller for the larger value of rotation parameter $a$.

\begin{figure*}
\begin{tabular}{|c|c|c|}
\hline
\includegraphics[width= 5.7 cm, height= 5 cm]{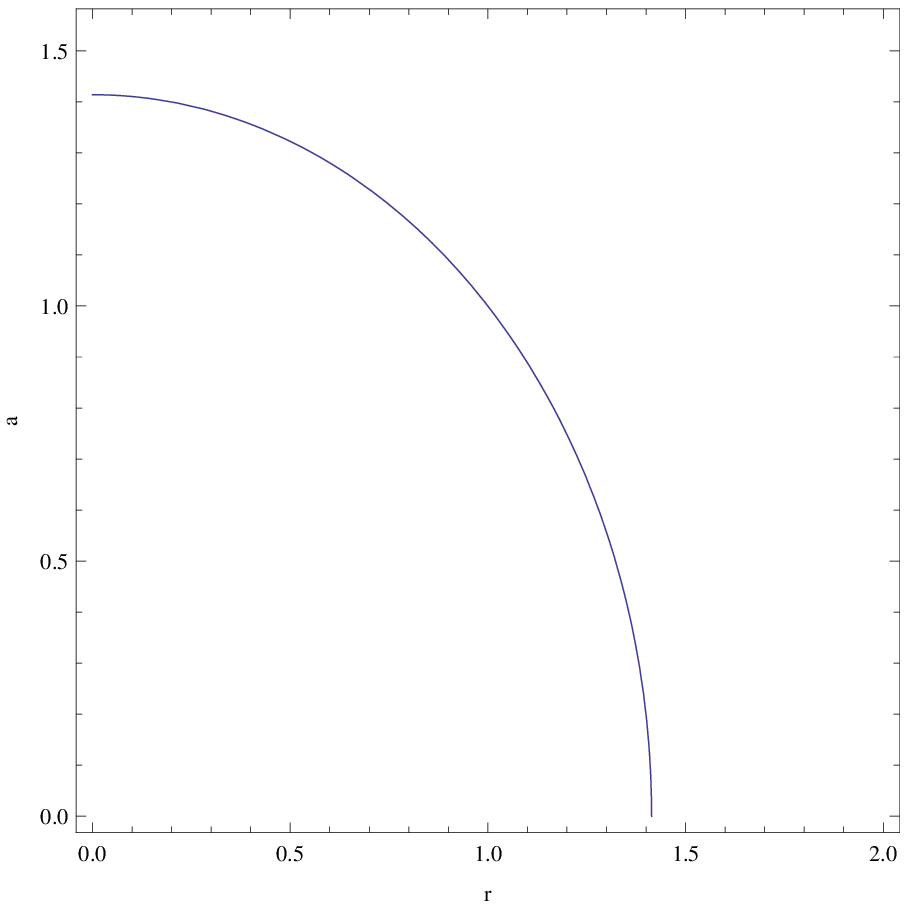}&
\includegraphics[width= 5.7 cm, height= 5 cm]{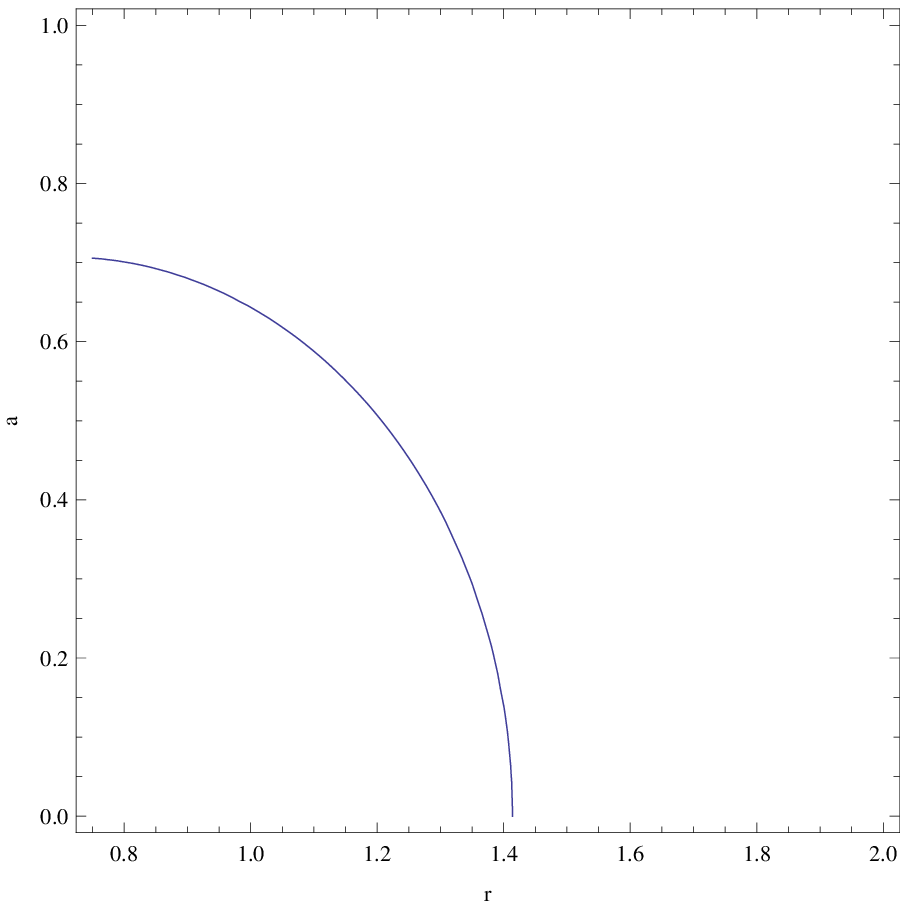}&
\\
\hline
\end{tabular}
\caption{\label{horizon}Plots showing the radial dependence of spin parameter $a$ and $b$ for the horizon. i) For spin parameter $a \neq b$. ii) For $a=b$.  }
\end{figure*}

\begin{figure*}
\begin{tabular}{|c|c|c|}
\hline
\includegraphics[width= 5.7 cm, height= 5 cm]{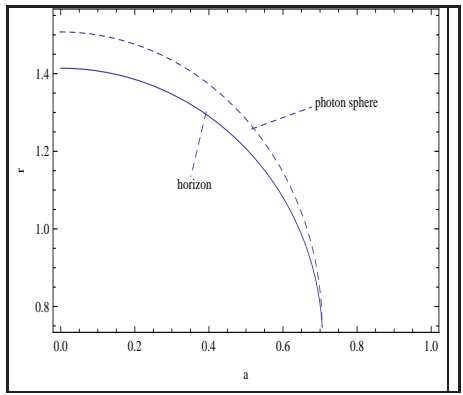}&
\\
\hline
\end{tabular}
\caption{\label{photon}Plots showing the radial dependence of spin parameter for the horizon and photon sphere for $a=b$. }
\end{figure*}

%For the metric (\ref{Kerr5D})\begin{equation}
%\sqrt{-g}=\frac{1}{2}\sin\theta \cos\theta \rho^2. \nonumber
%\end{equation}

%The expressions for the contravariant components of the metric are:
%\begin{eqnarray}
%&& g^{tt}=\frac{1}{\rho^2 } \left[(a^2-b^2)\sin^2\theta -
%\frac{(x+a^2)[\Delta+r_0^2(x+b^2)]}{\Delta} \right] , \nonumber \\
%&& g^{t\phi}=\frac{ar_0^2(x+b^2)}{\rho^2 \Delta} ,\hspace{1cm}
 %g^{t\psi}=\frac{br_0^2(x+a^2)}{\rho^2 \Delta} , \nonumber \\
 %&& g^{\phi\phi}=\frac{1}{\rho^2} \left[\frac{1}{
%\sin^2\theta}-\frac{(a^2-b^2)(x+b^2)+b^2r_0^2}{\Delta}\right], \nonumber \\
%&& g^{\psi\psi}=\frac{1}{\rho^2} \left[\frac{1}{
%\cos^2\theta}+\frac{(a^2-b^2)(x+a^2)-a^2r_0^2}{\Delta} \right] , \nonumber \\
%&& g^{\phi \psi}=-\frac{abr_0^2}{\rho^2 \Delta} ,\hspace{1cm}
% g^{xx}=4\frac{\Delta}{\rho^2} ,\hspace{1cm}
% g^{\theta \theta}=\frac{1}{\rho^2} .
%\end{eqnarray}
The metric (\ref{Kerr5D}) is invariant on taking the following transformation:
\begin{eqnarray}
a \leftrightarrow b, \;\; \theta\leftrightarrow \left(\frac{\pi}{2}-\theta\right), \;\; \phi\leftrightarrow\psi.
\end{eqnarray}This metric possesses three killing vectors $\partial_{t}$, $\partial_{\phi}$ and $\partial_{\psi}$. For $a=b$ the metric has two additional Killing vectors \cite{VD}.

In Fig.~\ref{horizon}, we have shown the possible range of rotation parameters for the case $a\neq b$ and $a=b$. When $a\neq b$, range of rotation parameter is $0<a<1.4$, on the other hand for $a=b$ it is $0<a<1/\sqrt{2}$.  In Fig.~\ref{photon}, we have shown the change in the behavior of photon sphere with horizon for $a=b$ case. It is interesting to note that with the increase in the value of rotation parameter  photon sphere is coming close to the central object, it is very important because photon sphere will show the shape of the shadow.

\section{Particle Motion}

To study the equation of motion of photons in the field of a 5D
rotating Myers-Perry black hole, we begin with the Lagrangian
which reads \begin{equation} L = \frac{1}{2} g_{\mu
\nu}\dot{x}^{\mu}\dot{x}^{\nu},
\end{equation}where an overdot denotes the partial derivative with respect to an affine parameter. Therefore, the momenta calculated for the metric (\ref{Kerr5D}) are:
\begin{eqnarray}
p_t &=& \left(-1+\frac{r_0^2}{\rho^2}\right)\dot{t}+\frac{r_0^2}{\rho^2}\dot{\phi}+\frac{r_0^2 b \cos^2\theta}{\rho^2}\dot{\psi}, \nonumber \\
p_{\phi}&=&\frac{r_0^2 a \sin^2\theta}{\rho^2}\dot{t}+\left(x+a^2+\frac{r_0^2 a^2 \sin^2\theta}{\rho^2}\right)\sin^2\theta \dot{\phi} \nonumber \\ && + \frac{r_0^2 a b \sin^2\theta \cos^2\theta}{\rho^2} \dot{\psi}, \nonumber \\
p_{\psi}&=& \frac{r_0^2 b \cos^2\theta}{\rho^2}\dot{t} + \frac{r_0^2 a b \sin^2\theta \cos^2\theta}{\rho^2} \dot{\phi} \nonumber \\ &&  + \left(x+b^2+\frac{r_0^2 b^2 \cos^2\theta}{\rho^2}\right)\cos^2\theta \dot{\psi}, \nonumber \\
p_x &=& \frac{\rho^2}{4\Delta}\dot{x}, \nonumber \\
p_\theta &=& \rho^2 \dot{\theta},
\end{eqnarray}where $p_t = -E$, $p_{\phi} = L_{\phi}$ and $p_{\psi}=L_{\psi}$ correspond to  energy and angular momentum with respect to the respective rotation axis  respectively.

\subsection{Photon orbits}

In order to analyze the general orbit of photon around a black
hole, we study the separability of the Hamilton-Jacobi equation
for which we adopt the approach originally suggested by Carter
\cite{Carter68}. It is straightforward to see that the
Hamilton-Jacobi equation in 5D black hole space-time
(\ref{Kerr5D}) with the metric tensor $g_{\mu\nu}$ takes the
following general form:
\begin{eqnarray}\label{Hamilton-Jacobieq}
-\frac{\partial S}{\partial \lambda} =
\frac{1}{2} g^{\mu \nu} \frac{\partial S}{\partial x^\mu}
\frac{\partial S}{\partial x^\nu},
\end{eqnarray}
with an affine parameter $\lambda$ and an action $S$ which can
be decomposed as a sum:
\begin{eqnarray}\label{Hamilton-Jacobiac}
S= \frac{1}{2} m^2 \lambda -Et+ L_{\phi}\phi + L_{\psi}\psi +
S_\theta (\theta) + S_x (x) \ ,
\end{eqnarray}
in order to separate the equation of motion.

For the photon the mass of the particle $m$ is equal to zero
($m=0$) and from (\ref{Hamilton-Jacobieq}) and
(\ref{Hamilton-Jacobiac}) we can conclude:
\begin{eqnarray}\label{thetaeq}
 \left( \frac{\partial S_\theta}{\partial \theta} \right)^2 -
E^2 (a^2 \cos^2\theta+ b^2 \sin^2\theta ) +\frac{L_{\phi}^2}
{\sin^2\theta} + \frac{L_{\psi}^2}{ \cos^2\theta} \nonumber \\  = {\cal K} ,
\end{eqnarray}
and
\begin{eqnarray}\label{xeq}
 4 \Delta \left(\frac{\partial S_x}{\partial x} \right)^2 - E^2 x -
\frac{r_0^2(x+a^2)(x+b^2)}{\Delta} \, \Sigma^2  \nonumber \\ -(a^2-b^2) \left(\frac{ L_{\phi}^2}{(x+a^2)}  -\frac{L_{\psi}^2}{(x+b^2)}\right)=-{\cal K}  ,
\end{eqnarray}
with ${\cal K}$ as a constant of separation and
\begin{equation}\label{EE0}
\Sigma= E+ \frac{a L_{\phi}}{x+a^2}+\frac{b L_{\psi}}{x+b^2}. \nonumber
\end{equation}
Equations (\ref{thetaeq}) and (\ref{xeq}) can be recast into the form:
\begin{eqnarray}\label{eqs}
\frac{\partial S_\theta}{\partial \theta}  = \sqrt{\Theta} , \nonumber
\hspace{1cm}
\frac{\partial S_x}{\partial x} =  \sqrt{X}, \nonumber
\end{eqnarray}where $X$ is given by \begin{eqnarray}\label{XX}
X= \frac{{R}}{4 \Delta^2}. \nonumber
\end{eqnarray}
\begin{figure*}
\begin{tabular}{|c|c|c|}
\hline
\includegraphics[width= 5.7 cm, height= 5 cm]{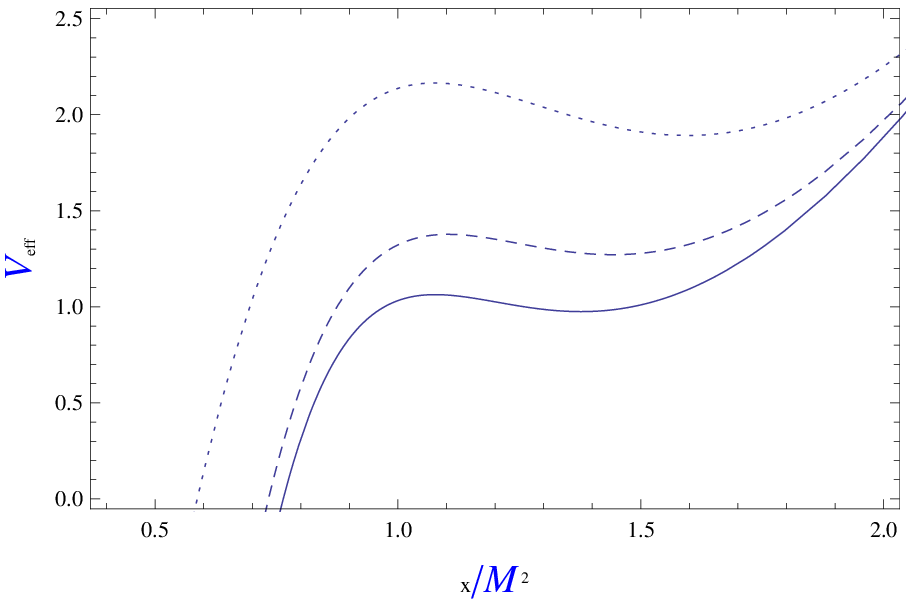}&
\includegraphics[width= 5.7 cm, height= 5 cm]{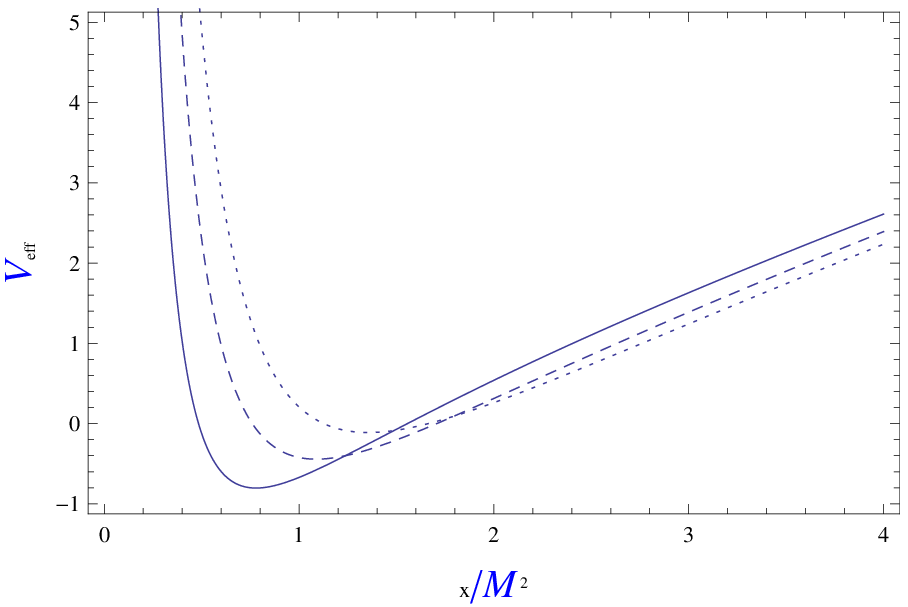}&
\\
\hline
\end{tabular}
\caption{\label{Veff}Plots showing the radial dependence of effective potential for rotation parameter $a$ and $b$. (Left) For different values of $a$: $a=1.2$ (solid line), $a=1$ (dashed line), $a=0.9$ (dot dashed line) and $b=0$. (Right) For different values of $a=b$: $a=0.4$ (solid line), $a=0.3$ (dashed line), $a=0.2$ (dot dashed line). }
\end{figure*}
Here functions $\Theta$ and $R$ are given by
\begin{eqnarray}\label{Theta}
\Theta =E^2 (a^2 \cos^2\theta+ b^2 \sin^2\theta )
-\frac{\cos^2\theta}{\sin^2\theta}L_{\phi}^2- \frac{\sin^2\theta}{\cos^2\theta}L_{\psi}^2 \nonumber \\ + {\cal K}  ,
\end{eqnarray}
 \begin{eqnarray}\label{X}
{R}&=& \Delta \Big[ x E^2
 +(a^2-b^2)\left(\frac{L_{\phi}^2}{(x+a^2)}
  -\frac{L_{\psi}^2}{(x+b^2)}\right) \nonumber \\&& -{\cal K}\Big]
+r_0^2(x+a^2)(x+b^2) \Sigma^2.
\end{eqnarray}
 We can write the Hamilton-Jacobi action in terms of these
functions:
\begin{eqnarray}\label{Hamilton-Jacobi}
S=\frac{1}{2} m^2 \lambda -E t+ L_{\phi}\phi + L_{\psi}\psi + \int^\theta
\sqrt{\Theta} d\theta +\int^x \sqrt{X}dx .
\end{eqnarray}

Hence according
to~\cite{VD,claus}
the null
geodesic
equations
can be
obtained
from the
Hamilton-Jacobi
equation as
\begin{eqnarray}\label{EOM}
\rho^2 \dot{\theta} &=& \sqrt{\Theta} , \nonumber \\
\rho^2 \dot{x} &=& 2 \sqrt{{R}} ,\nonumber \\
\rho^2 \dot{t} &=& E \rho^2 + \frac{r_0^2(x+a^2)(x+b^2)}{\Delta} \Sigma ,\nonumber \\
\rho^2 \dot{\phi} &=&  \frac{L_{\phi}}{\sin^2 \theta}-\frac{a r_0^2(x+b^2)}{\Delta} \Sigma-\frac{(a^2-b^2)L_{\phi}}{(x+a^2)}, \nonumber \\
\rho^2 \dot{\psi} &=& \frac{L_{\psi}}{\cos^2 \theta}-\frac{br_0^2(x+a^2)}{\Delta}
\Sigma+\frac{(a^2-b^2)L_{\psi}}{(x+b^2)}.
\end{eqnarray}

These equations governs the propagation of light in the spacetime of 5D black hole.
In order to obtain the boundary of the shadow of the black hole, we need to study the radial motion of photons. First, we rewrite the radial equation as\begin{eqnarray}
\rho^2 \dot{r} &=& \sqrt{R}.
\end{eqnarray}The effective potential reads \begin{eqnarray}\label{veff}
V_{eff}&=&\frac{1}{2}\frac{\Delta}{\rho^2} \Big[ x E^2
 +(a^2-b^2)\left[\frac{L_{\phi}^2}{(x+a^2)}
  -\frac{L_{\psi}^2}{(x+b^2)}\right] \nonumber \\&& -{\cal K}\Big]
+r_0^2(x+a^2)(x+b^2) \Sigma^2.
\end{eqnarray}
One can think of an effective potential for the photon, which has a barrier with a maximum, goes to negative infinity below the horizon, and asymptotes to zero
at $x \rightarrow \infty$.  In the simplest case of Schwarzschild case, the effective potential for photons has a maximum at $3M$, the location of the unstable orbit (there is no minimum of this potential). For rotating case, the picture is qualitatively the same, but a little more complex, because the spin breaks the spherical symmetry of the system. The general behavior of effective potential for the 5D Myers-Perry black as a function of $x$ for different values of rotation parameter is shown in Fig. \ref{Veff}. For  $b=0$ and varying $a$, the effective potential shows one maximum and one minimum which corresponds to the unstable and stable circular orbits, respectively. It is seen that with the increase in the value of rotation parameter peak of the graph shifts to left which signifies that the circular orbits are shifting towards the central object. For $a=b$, it is observed that there is only one minimum which shows the presence of stable circular orbits.

Now, we know that the equation of motion defined in Eq.
(\ref{EOM})  determines the propagation of light in the space time
of 5D rotating black hole. According to~\cite{Bambi} the
apparent silhouette of the 5D rotating black hole in principle can
be obtained through the analysis of the photon orbits around the
5D rotating black hole. In the axial-symmetric and stationary 5D
space-time three impact parameters expressed in terms of constants
of motion $E$, $L_{\phi}$ and $L_{\psi}$ and the constant of
separability ${\cal K}$ play the role of the characteristics of
each photon orbit. Combining these quantities, one can define the
following impact parameters $\xi_{1}=L_{\phi}/E$,
$\xi_{2}=L_{\psi}/E$ and $\eta={\cal K}/E^2$ for photon orbits
around the general axial-symmetric stationary 5D rotating black
hole. Henceforth, for simplicity of analysis equal values of
rotation parameters $(a=b)$, we assume $a=b$, which gives
\begin{eqnarray} R &=& \Big[(x+a^2)^2-r_0^2\Big] \Big( x E^2
-{\cal K}\Big) +r_0^2(x+a^2)^2 \nonumber \\ && \times \Big[E
+\frac{a}{x+a^2}(L_{\phi}+L_{\psi})\Big]^2 .
\end{eqnarray}
\begin{figure*}
\includegraphics[width= 8 cm, height= 8 cm]{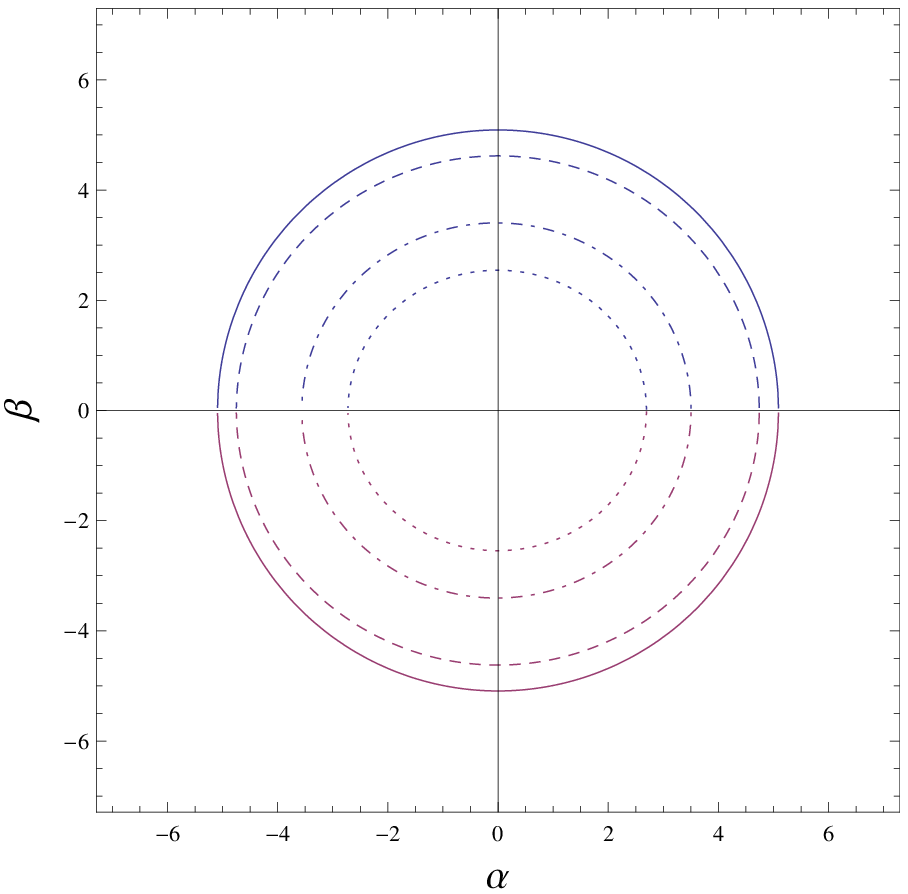}
\includegraphics[width= 8 cm, height= 8 cm]{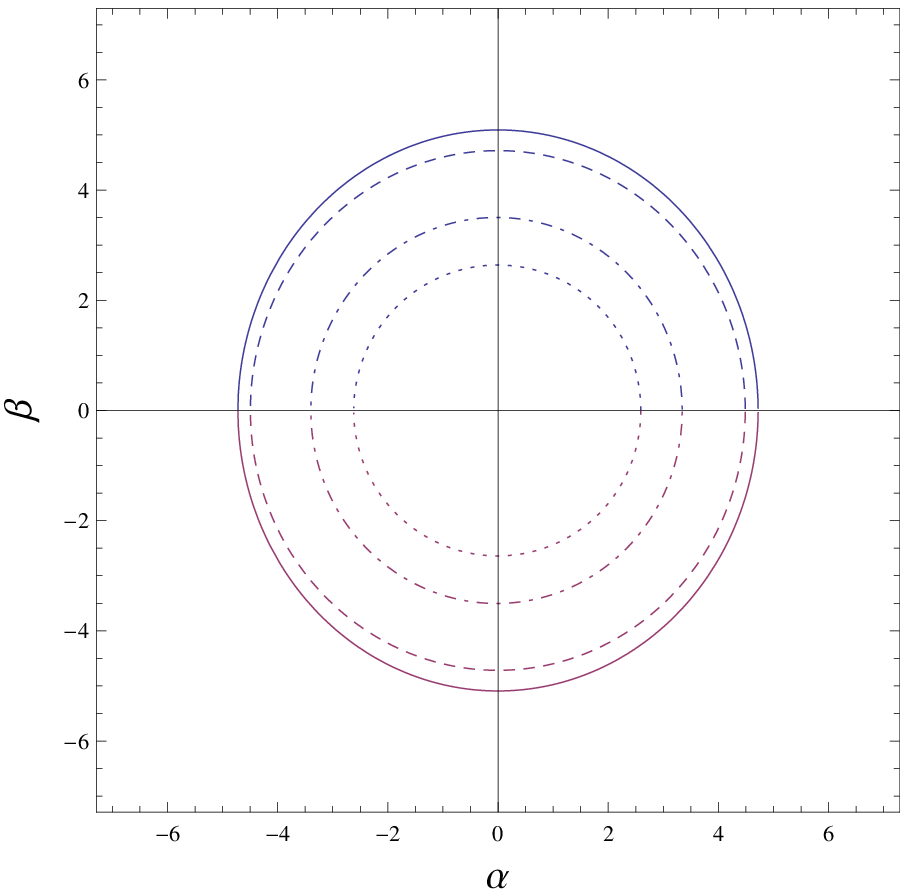}
\\
\includegraphics[width= 8 cm, height= 8 cm]{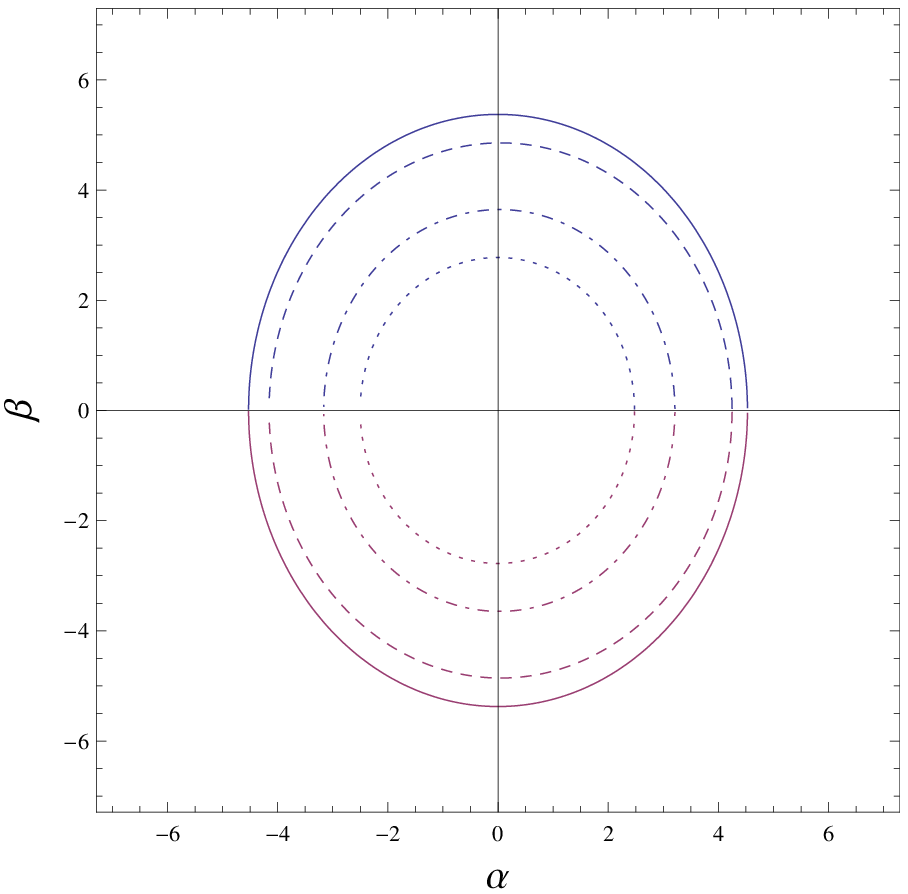}
\includegraphics[width= 8 cm, height= 8 cm]{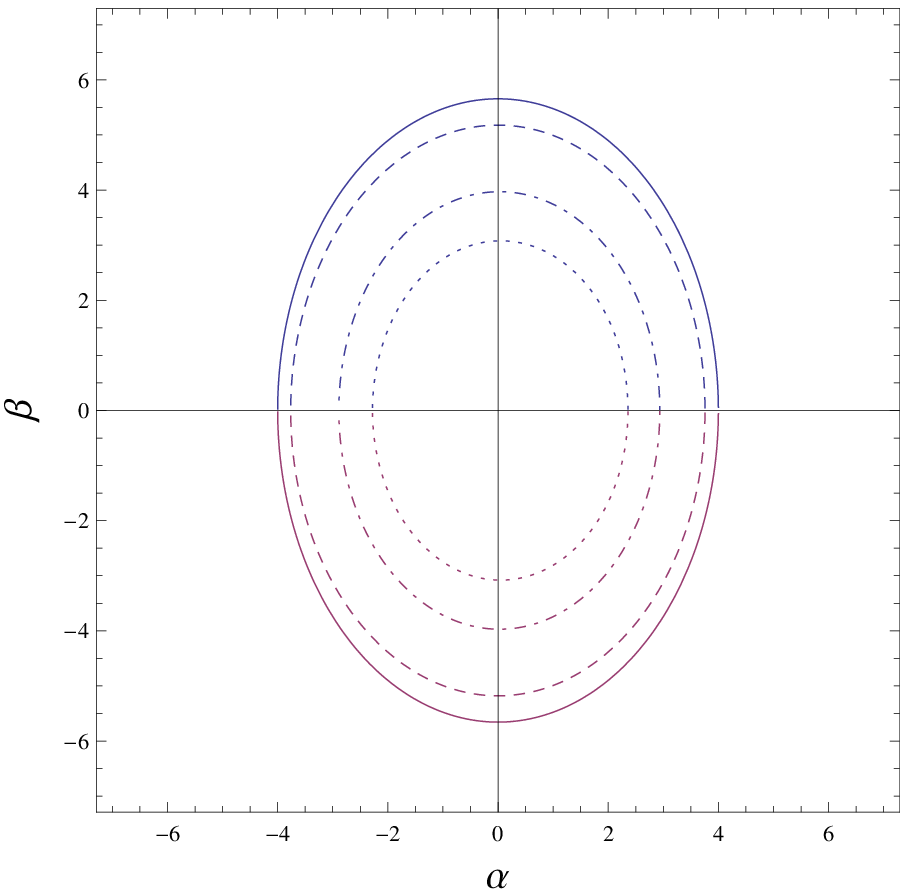}
\caption{\label{Shadow}Shadow cast by 5D rotating black hole taking $a=b$. Top: For $\theta_0=0$ (left) \; for $\theta_0=\pi/6$ (right). Bottom: For $\theta_0=\pi/4$ (left)\; $\theta_0=\pi/3$ (right). $\theta_0=0$ is same with  $\theta_0=\pi/2$ and  $a$ is as following: $a=0$ (solid line), $a=0.4$ (dashed line), $a=0.5$ (dot-dashed line), $a=0.6$ (dotted line).}
\end{figure*}

According to~\cite{Bambi} the unstable photon orbits are
responsible for an apparent silhouette of the 5D rotating black
hole. Consequently the boundary of shadow of the 5D rotating black
holes can be defined through the conditions
\begin{equation}\label{condition}
V_{eff}=0, \partial V_{eff}/\partial r=0 (\textrm{or}
R(r)=0=\partial R(r)/\partial r)\ .
\end{equation}
 Then using the Eq.~(\ref{Veff}) one can obtain the contour of the
shadow as
\begin{widetext}
\begin{eqnarray}
\eta(r)= \frac{a^6-5 x^2+2 x^3 +a^4(1+5x)+a^2 x (7x-4)-2(a^4+2 a^2 x+x(x-2))\sqrt{a^2+x}}{2 (x+a^2-1)^2},
\end{eqnarray}
\begin{eqnarray}
\xi_1(r)+\xi_2(r) = \frac{-a^4+2a^2(x-1)+x^2+(a^4+2a^2 x+x(x-2))\sqrt{a^2+x}}{2a(x+a^2-1)}.
\end{eqnarray}
\end{widetext}

%\begin{figure}
%\includegraphics[width=0.9\linewidth]{scheme.eps}
%\caption{\label{DsRs} The observables for the apparent shape of the 5D black holes are the radius $R_s$ and the distortion parameter $\delta_s=D_{cs}/R_s$, where $D_{cs}$ is the %difference  between the left endpoints of the circle and of the shadow.}
%\end{figure}
\begin{figure*}
\begin{tabular}{|c|c|c|}
\hline
\includegraphics[width=0.45\linewidth]{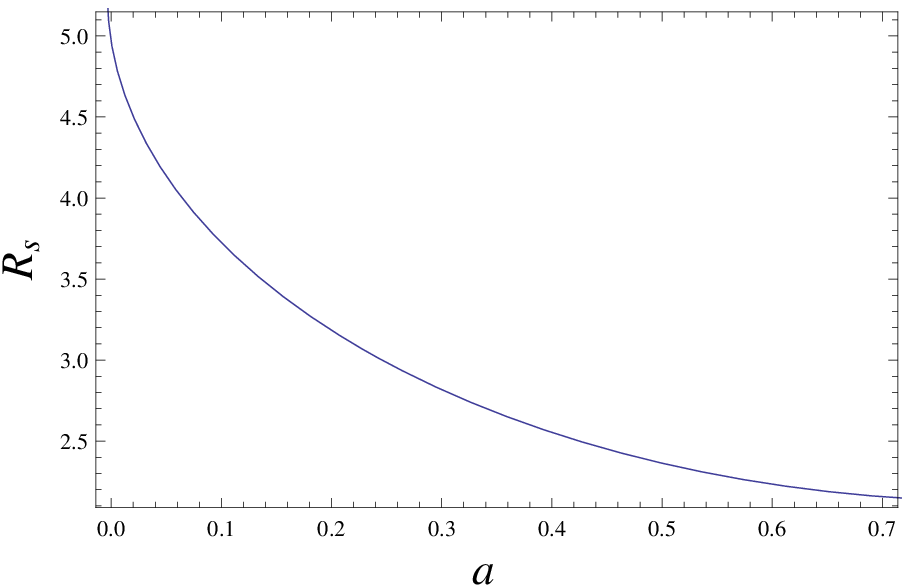}&
\includegraphics[width=0.45\linewidth]{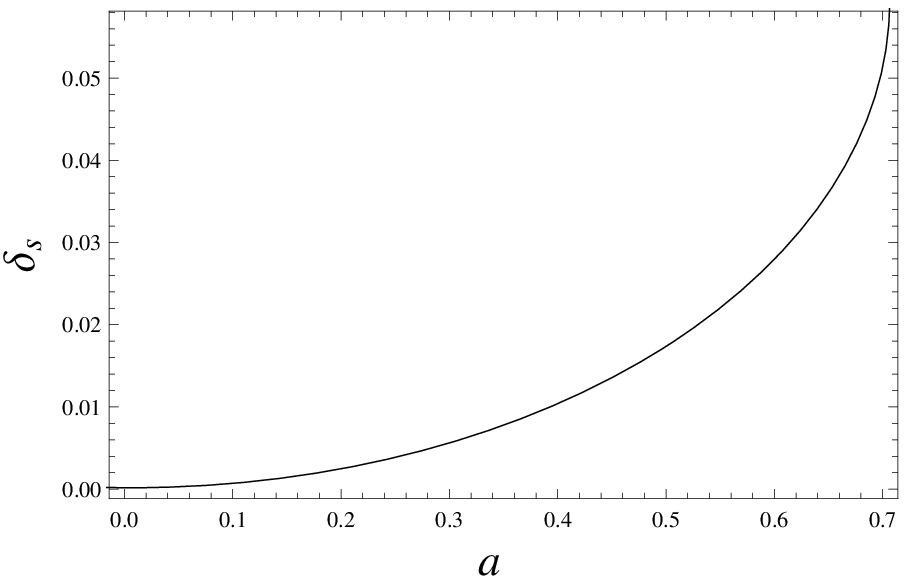}
\\
\hline
\end{tabular}
\caption{\label{Rs,Delta}Plots showing the dependence of the observables, the radius of the shadow $R_s$ (left) and distortion parameter $\delta_s$ (right) on the rotation parameter with inclination angle $\theta_0=0$.}
\end{figure*}
\begin{widetext}
For $\theta=0$, $L_{\phi}=0$ which implies $\xi_2(r)=0$, therefore
\begin{equation}
\xi_1(r) = \frac{-a^4+2a^2(x-1)+x^2+(a^4+2a^2 x+x(x-2))\sqrt{a^2+x}}{2a(x+a^2-1)},
\end{equation}
and for $\theta=\pi/2$, $L_{\psi}=0$ which implies $\xi_1(r)=0$, thus
\begin{equation}
\xi_2(r) = \frac{-a^4+2a^2(x-1)+x^2+(a^4+2a^2 x+x(x-2))\sqrt{a^2+x}}{2a(x+a^2-1)}.
\end{equation}
\end{widetext}
\begin{figure*}
\begin{tabular}{|c|c|c|c|c|}
\hline
\includegraphics[width= 8 cm, height= 6 cm]{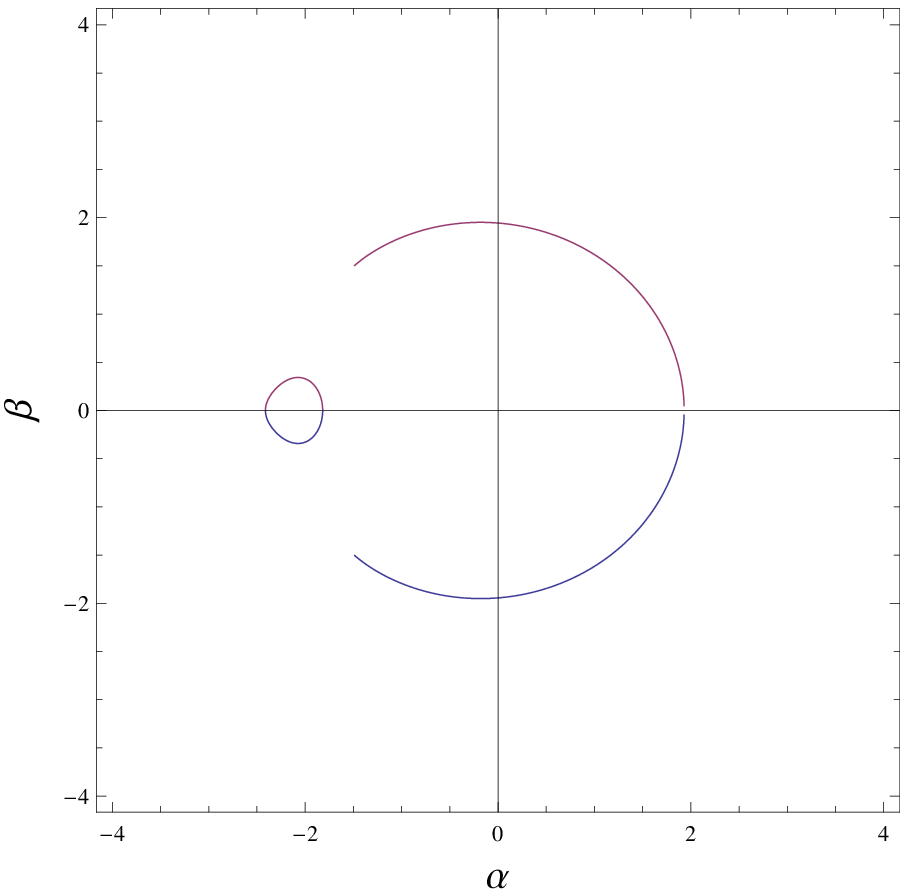}&
\includegraphics[width= 8 cm, height= 6 cm]{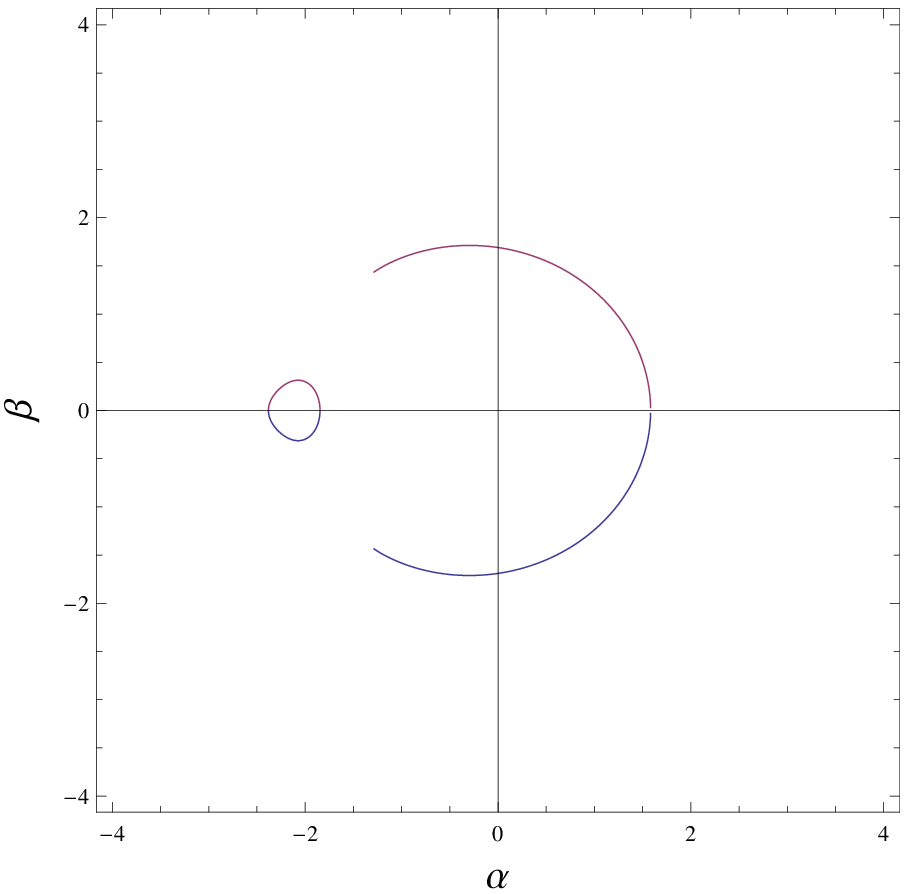}
\\ \hline
\hline
\includegraphics[width= 8 cm, height= 6 cm]{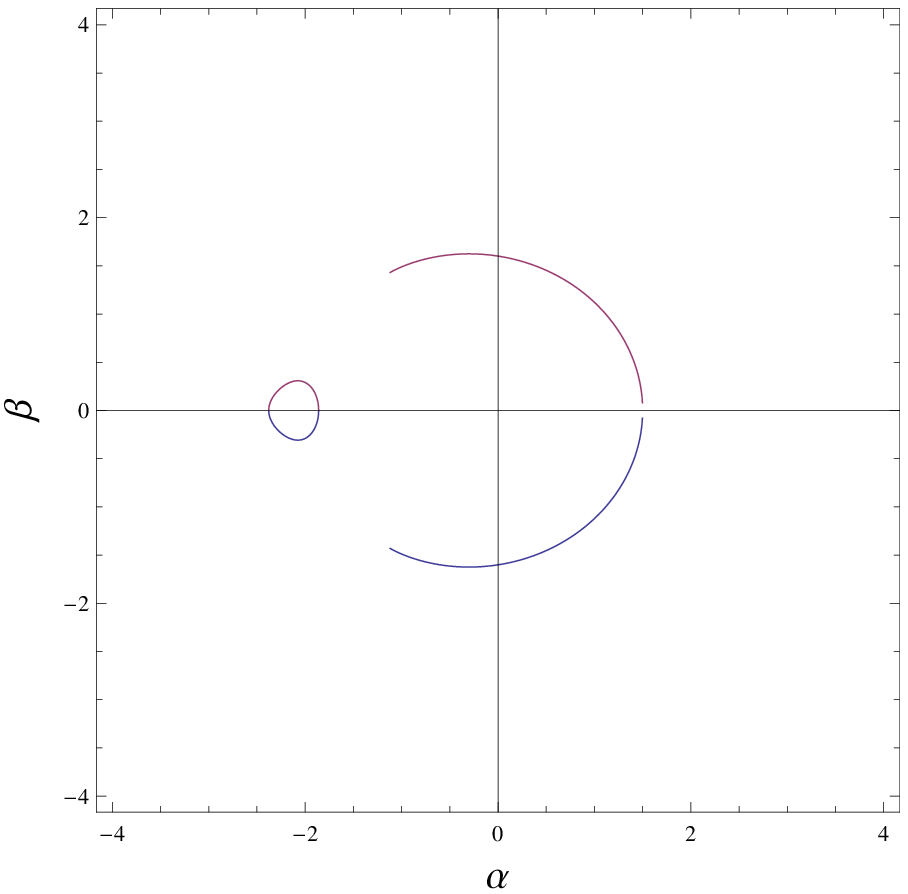}&
\includegraphics[width= 8 cm, height= 6 cm]{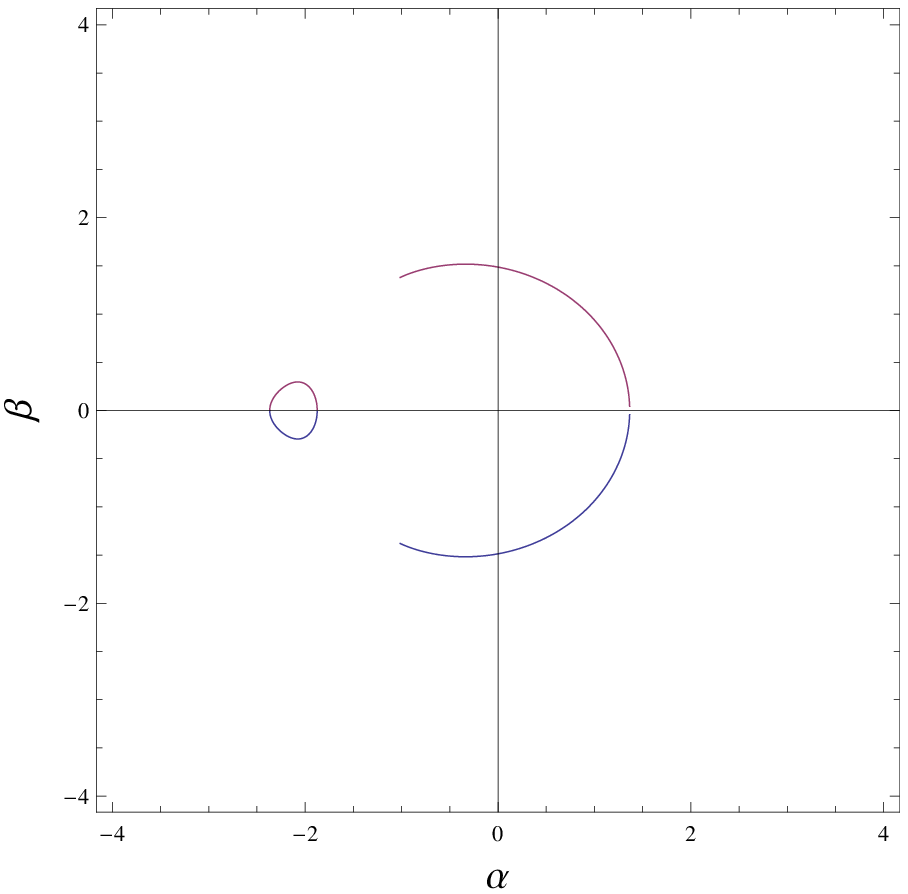}
\\
\hline
\end{tabular}
\caption{\label{NS}Shadow cast by naked singularity  taking $a=b$
and for the different value of rotation parameter $a$:  $a=0.9$
(top panel, left), $a=1$ (top panel, right), $a=1.1$ (bottom,
left) and $a=1.2$ (bottom, right). Here inclination angle is
$\theta=0$ and it is same with $\theta=\pi/2$.}
\end{figure*}

\section{{Shadow of a 5D rotating black hole}}

Here we investigate the shadow of a black hole and naked
singularity. It is possible to obtain equatorial orbits of photon
around 5D black hole from the effective potential. The photon
orbits are in general of three types: scattering, falling and
unstable ones which can be treated in terms of effective
potential in the following way~\cite{Bambi}: (i)  the photons
arriving from infinity with energy being larger than the barrier
of the effective potential penetrate the horizon and fall down
into the black hole along \textit{the falling orbits}, (ii)
 the photons arriving from infinity with
energy less than the barrier of the effective potential move along
\textit{the scattering orbits} and comes back to infinity, and
(iii)  the maximum of the effective potential separates the
captured and the scattering orbits and defines unstable orbits of
constant radius (which is circle located at $r=3M$ for the
Schwarzschild black hole) which is responsible for the apparent
silhouette of the 5D rotating black hole. An observer far away
from the black hole will be able to see only the photons scattered
away from the black hole, while those captured by the black hole
will form a dark region. If 5D rotating black hole appears between a
light source and a distant observer, the photons with small impact
parameters fall into the black hole and form a dark zone in the
sky which is usually called as black hole shadow.

The shadow of a black hole is better described by the celestial
coordinates $\alpha$ and $\beta$ \cite{Vazquez}, which in 5D case
are modified to:
\begin{equation}  \label{5Dalpha}
\alpha=\lim_{\tilde{r}_{0}\rightarrow \infty} -\tilde{r}_{0}^{2}\left(\sin\theta_{0}\frac{d\varphi}{dr}+\cos\theta_{0}\frac{d\psi}{dr}\right),
\end{equation}
and
\begin{equation}  \label{5Dbeta}
\beta=\lim_{\tilde{r}_{0}\rightarrow \infty}\tilde{r}_{0}^{2}\frac{d\theta_0}{dr}.
\end{equation}
Here according to the standard  definitions $\tilde{r}_{0}$ is
the distance from the black hole to observer, the coordinate
$\alpha$ is
the apparent
perpendicular
distance between the image and the axis of symmetry, and the
coordinate $\beta$ is the
apparent
perpendicular distance between
the image and its projection on the equatorial plane. For further
details and a useful diagram, see Ref.~\cite{Vazquez}.

Next we consider that the observer is far away from the black hole
and hence,  $r_0$ tends to infinity, and the angular
coordinate of the distant observer $\theta_0$ is the inclination
angle between the line of sight of the distant observer and the
axis of rotation of the central gravitating object. In order to
analyze the shape of the 5D black hole two coordinates $(\alpha,
\beta)$ are introduced. In order to derive Eqs.(\ref{5Dalpha}) and
(\ref{5Dbeta}) one can write $(\alpha, \beta)$  in the Euclidean
coordinates system, then transform it to spherical coordinates,
and apply the geometrical formalism of the straight line
connecting the apparent position of the image to  the position of
the distant observer.

On using Eq. (\ref{EOM}), in Eqs. (\ref{5Dalpha}) and (\ref{5Dbeta}), and taking the limit of a far away observer, we found the celestial coordinates, as a function of the constant of motion, take the form
\begin{eqnarray}\label{AB}
\alpha &=& -\left( \xi_{1}\frac{1}{\sin\theta_0} + \xi_{2}\frac{1}{\cos\theta_0} \right),\nonumber \\
\beta &=& \pm \sqrt{\eta - \xi_{1}^2 \cot^2\theta_0 - \xi_{2}^2 \tan^2\theta_0 + a^2}.
\end{eqnarray}In the equatorial plane $\theta_0=\pi/2$, then the celestial coordinates can be written as \begin{eqnarray}
\alpha &=& -\xi_1, \nonumber \\ \beta &=& \pm \sqrt{\eta+a^2}.
\end{eqnarray}For $\theta=0$, \begin{eqnarray}
\alpha &=& -\xi_2, \nonumber \\ \beta &=& \pm \sqrt{\eta+a^2}.
\end{eqnarray}
%\begin{eqnarray}
%\alpha^2 + \beta^2 &=& \frac{1}{a^2(a^4-r^4)^2}\Big(r^{12}+2(7a^2-4)r^{10}+2(17a^2\nonumber \\ &&-20a^2+8)r^8  +4a^4(7a^2-6)r^6)+a^6(9a^2-8)r^4\nonumber \\ &&+6a^{10}
%r^2+4a^{12}\Big)
%\end{eqnarray}
Now we show the shadows of 5D rotating black hole using Eq. (\ref{AB}). In Fig. \ref{Shadow}, we plot $\alpha$ Vs $\beta$ to show the contour of the shadow of the black holes for different values of rotation parameter at different inclination angles. It can be seen from the figure that with increasing the value of rotation parameter the effective size of the shadow is decreasing.

\begin{figure}
\begin{tabular}{|c|c|c|}
\hline
\includegraphics[width=0.80\linewidth]{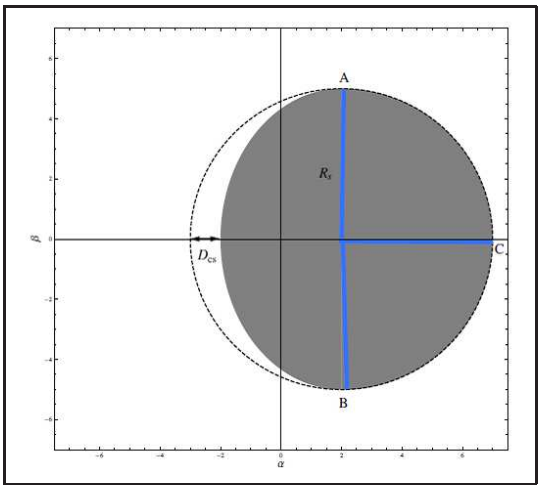}
\\
\hline
\end{tabular}
\caption{\label{ABC} The observables parameters the radius $R_{s}$ and the distortion
parameter $\delta_{s}=D_{cs}/R_{s}$ are described as the apparent shape
of black hole(adopted from ~\cite{Hioki08,Abdujabbarov}).}
\end{figure}

In order to observe the shadow of a black hole, it is useful to
introduce two observables the radius $R_s$ and the distortion
parameter $\delta_s$ that approximately characterizes its shape. As
shown in Fig.~\ref{ABC}, in Ref \cite{Abdujabbarov,Hioki08}, the shape of the shadow of the black
hole is a circle, so we approximate the shadow by a circle passing
through three points, which are located at the top position $(A)$,
the bottom position $(B)$, and the most left end of its boundary
$(C)$ defined by the coordinates $(\alpha_t, \beta_t)$, $(\alpha_b,
\beta_b)$ and $(\alpha_r, 0)$ respectively. The radius $R_s$ of the
shadow is hereby defined by the radius of this circle. The point $C$
corresponds to the unstable retrograde circular orbit when seen from
an observer on the equatorial plane. The distortion parameter
$\delta_s$ is defined as
\begin{equation}
\delta_{s}=\frac{D_{cs}}{R_s},
\end{equation}
where $D_{cs}$ is the difference between the right endpoints of the circle and of the shadow.

The observable $R_s$ is defined as
\begin{equation}
R_s = \frac{(\alpha_t - \alpha_r)^2 + \beta_t^2}{2(\alpha_t - \alpha_r)}.
\end{equation}

In Fig.~\ref{Rs,Delta}, the observables $R_s$ and $\delta_s$  are
plotted as a function of the rotation parameter $a$.  The
observable $R_s$ is the main quantity being responsible for the
size of the 5D black hole shadow. From Fig.~\ref{Rs,Delta}, we
observe that the observable $R_s$ monotonically decrease with the
increase in the value of rotation parameter . Whereas  the
observable $\delta_s$ gives distortion of the shadow with respect
to the circumference of reference, which increases with $a$.

In Fig. \ref{Shadow}, the shadows are plotted for different values of the rotation parameter $a$ and the angle of inclination $\theta_0$, for
the $5D$ Myers-Perry black hole. In all the cases, we take values for the angular momentum that go from 0 to 0.6, and it can be seen clearly that the shape of the shadow is decreasing with the increase in the value of rotation parameter. Also, the shape is more distorted with the increase in the value of rotation parameter. It is interesting to note that the rate of distortion also depends on the angle of inclination, it is seen that rate of distortion is more for $\theta_0=\pi/4$ than in the case of $\theta_0=0$.

In the case of Schwarzschild black hole, the apparent image of the black hole is a perfect circle of radius $\approx 5.20 M$ \cite{Bambi}. The main feature of the shape of Kerr black holes is the asymmetry along the spin axis, because of the different effective potential for photons orbiting around the black hole in one or the other direction. The radius of the unstable circular orbit is smaller for photons with angular momentum parallel to the black hole spin and that slightly flattens the black hole shadow on one side.

 Whereas, in the Brane-world case \cite{Amarilla1}, in addition to the angular momentum, the tidal charge term also deforms the shape of the shadow. For a given value of the rotation parameter, the presence of a negative tidal charge enlarges the shadow and reduces its deformation with respect to Kerr spacetime, while for a positive charge, the opposite effect is obtained.  However, in the Brane world, for fixed tidal charge,  there is  a very small variation in the size of the shadow as a function of $ a $.  In a similar study in for the Kaluza-Klein black holes in Einstein gravity coupled to a Maxwell field and a dilaton, the size and the shape of the shadow depend on the mass, the charge, and the angular momentum and, for fixed values of these parameters, the shadow is slightly larger and less deformed than for its Kerr-Newman counterpart.  On the contrary, our results demonstrate that the size of the the $5D$ Myers-Perry black hole shadow is smaller than for Kerr black hole shadow.   In general, the size of the shadow decreases with rotation parameter in 5D Myers-Perry black hole compared to the four-dimensional Kerr black hole

It is well known that, at high energy the absorption cross section of a black hole oscillates around a limiting constant value and the black hole shadow corresponds to its high energy absorption cross section for the observer located at infinity. The limiting constant value has been given in terms of geodesics, and is also analyzed for wave theories. For a black hole endowed with a photon sphere, the limiting constant value is same as
the geometrical cross section of this photon sphere \cite{Misner}.
\begin{figure}
\begin{tabular}{|c|c|c|}
\hline
\includegraphics[width=0.80\linewidth]{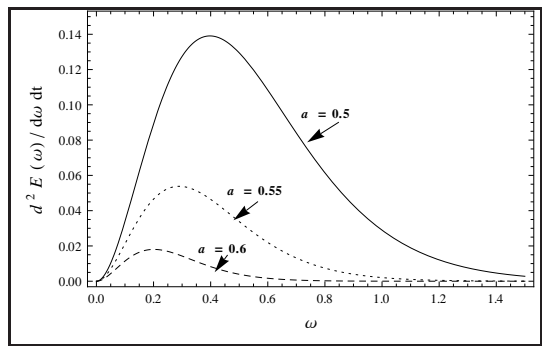}
\\
\hline
\end{tabular}
\caption{\label{energy}Plots showing the rate of energy emission varying with the frequency for different values of rotation parameter.}
\end{figure}
Here, we compute the energy emission rate using \begin{equation}
\frac{d^2E(\omega)}{d\omega dt}= \frac{2 \pi^2 \sigma_{lim}}{\exp{\omega/T}-1}\omega^3,
\end{equation}where $T=(r_{+}^2-a^2)r_{+} r_0^2/2\pi (r_{+}^2+a^2)^3$ is the Hawking temperature and $\sigma_{lim}$ is the limiting constant value for a spherically symmetric black hole around which the absorption cross section oscillates. The limiting constant value can be expressed as \begin{equation}
\sigma_{lim} \approx \pi R_s^2. \nonumber
\end{equation}Hence
\begin{eqnarray}
\frac{d^2E(\omega)}{d\omega dt}=\frac{2\pi^3 R_s^2}{e^{\omega/T}-1}\omega^3. \nonumber
\end{eqnarray}
In Fig.~\ref{energy}, we plot the energy emission rate versus frequency for different value of rotation parameter and it is seen that with the increase in the values of rotation parameter the peak of energy emission rate decreases.

\subsection{Naked Singularity Shadow}

A naked singularity is defined as a spacetime singularity without
an event horizon which can be  seen by some observer. According to
the cosmic censorship conjecture  \cite{rp}, the singularities
that appear in gravitational collapse are always surrounded by an
event horizon. Moreover, according to the strong version of the
conjecture, such singularities are not even locally naked, i.e.,
no nonspacelike curve can emerge from such singularities (see
\cite{r1} for reviews on the cosmic censorship conjecture).   The
cosmic censorship conjecture has as yet no precise mathematical
formulation or proof for either version. Hence the cosmic
censorship conjecture remains one of the most important unsolved
problems in general relativity and gravitation theory today. But
as we are still far from having a general proof of this hypothesis
so it is interesting to study about the shadow of a naked
singularity.   Consequently, such a study is an important tool to
get insight to this unresolved issue. For example, ultrahigh
energy collisions and optical phenomena around Kerr naked
singularities and superspinars are studied, e.g.,
in~\cite{stuchlik14,stuchlik13,stuchlik12a,stuchlik12b,Patil11,Patil12}.

When, $r_0^2<4a^2$ or $a > 1/\sqrt{2}$, from Eq.~(\ref{horizon1})
one see that horizons do not exist and singularity is exposed to
an external observer, i.e., there exists a naked singularity
violating cosmic censorship conjecture.  For non-rotating case
($a=0$), one obtains a point naked singularity whilst the rotating
case ($a\neq0$) corresponds to a ring type naked singularity.  For
the visualization of naked singularities shadow, we show a contour
of them for different values of rotation parameters. In the case
of a Kerr naked singularity, the shadow consists of two parts: the
arc and the dark spot or the straight line. Next we show the
shadows of a 5D Myers-Perry naked singularity in Fig.~\ref{NS}. In
the case of a naked singularity, the event horizon does not exist
and then the apparent shapes drastically different from those of a
black hole. The unstable spherical photon orbit with a positive
radius  constructs an arc rather than a closed curve. This is
because the photons near the arc may come back to the observer due
to the nonexistence of horizon.  In Fig.~\ref{NS}, we have plotted
$\alpha$ vs $\beta$ to show the shadow of  naked singularities of
$5D$ rotating Myers-Perry spacetime. Indeed, the shadow consists
of two parts an arc and a dark spots.  Fig.~\ref{NS}, we observe
that arc of shadow tends to open with increase rotation parameter
$a$.  Whereas the size of shadow decreases with increasing black
hole's spin $a$.  Thus, we observed that the shadow of a  5D
rotating naked singularity, which consists of a dark arc and a
dark spot, is very different from a 5D rotating black hole one.
So, the two observables, viz. $R_s$ and $\delta_s$ defined above
are no longer valid for a naked singularity.  However, another two
new observables \cite{Hioki08,Wei} can be defined to describe the
shadow of a naked singularity.

\section{Conclusion}

The shadow of a black hole has been active subject of research in
the last few years because that the observations of a black hole,
in the centre of a galaxy, may be obtained in the near
future~\cite{falcke}. Therefore, the investigation of the shadow
may be very useful tool to study the nature of the black hole. A
black hole casts a shadow due to the strong gravitational field
effects near the black hole. In the case of four-dimensional Kerr
BH,  the apparent shape of the shadow is distorted mainly by its
spin parameter and the inclination angle.  We have extended
previous study, of shadow cast by a black hole, to 5D Myers-Perry
spacetime for the case of two equal rotation parameters.  There is
intriguing dissimilarity of this problem with the case of
four-dimensional Kerr black hole, e.g., while in the
four-dimensional Kerr case there exists stable circular orbit
around the black hole, they are absent for 5D Myers-Perry black
hole. Here we analyze the unstable spherical photon orbits 5D
Myers-Perry black hole.

We have investigated how the size and apparent shape of the black hole is distorted due to the extra dimension by analyzing unstable circular orbits, i.e., we have studied the effect of rotation parameter on the shape of the shadow of a 5D black hole.  We adopt two observables, the radius $R_s$ and the distortion parameter $\delta_s$, characterizing the apparent shape, we found that the shape of the shadow is affected by the value of rotation parameter.
Interestingly, the size of the shadow decreases with rotation parameter in 5D Myers-Perry black hole resulting in a smaller shadow than in the four-dimensional Kerr black hole.  Thus the larger value of rotation parameter $a$ leads to decrease in the size of shadow.  This may be understood by the fact that as the black hole starts spinning rapidly, it forces photons orbits to come closer resulting in the decrease in gravitational force acting on photons.  The deformation of the shadow is characterized by the observable $\delta_s$ which increases monotonically with rotation parameter $a$ and takes maximal value when black hole approaches extremal.  Thus,  the distortion also increases with the increase in rotation parameter. It corresponds to the deviation of the shape of the shadow from circular orbit and it is also seen that the rate of change of distortion with rotation parameter also depends on the angle of inclination. We have also found that, on the contrary to the Brane world where the size of shadow does not changes much for fixed tidal charged as a function of rotation parameter, the size of 5D Myers Perry black hole shadow decreases significantly.

We also extended our analysis to study the shadow of 5D
Myers-Perry naked singularity,  which have two parts the dark arc
and distorted circular shape. The results are similar to the Kerr
/ Kerr Newman naked singularities \cite{Hioki08}, where the shadow
has two parts, the dark arc and the dark spots. We also observe
the deviation of the peak of effective potential towards the
central object.

It will be of interest to  generalize these results to  the case of unequal rotation parameter. This and related study will be subject of our forthcoming paper.

\acknowledgements Two of the Authors (UP, SGG) would like to thank
University Grant Commission (UGC) for major research project grant NO. F-39-459/2010(SR) and to Institute of Nuclear Physics, Tashkent for kind hospitality while this work was being done. FA and BA's research is supported in part by the project F2-FA-F113 of the UzAS, by the Volkswagen Stiftung (Grant 86 866) and by the  ICTP through the OEA-NET-76, OEA-PRJ-29 projects. We would like to thank the anonymous referee for useful comments.


\begin{references}
\bibitem {HM} K.~Hioki and K.~I.~Maeda, Phys. Rev. D {\bf 80}, 024042 (2009).
\bibitem {BR} B.~C.~Bromley, F.~ Melia and S.~ Liu, Astrophys.J. {\bf 555}, L83 (2001); A.~deVries, Phys. Unserer Zeit {\bf 35}, 128 (2004).
\bibitem {Vries} A.~ de Vries, Class. Quant. Grav. {\bf 17}, 123 (2000).
\bibitem {Takahashi04} R.~ Takahashi, Astrophys. J. {\bf 611}, 996 (2004).
\bibitem {Bambi} C.~ Bambi and K.~ Freese, Phys. Rev.D {\bf 79}, 043002 (2009); C.~ Bambi and N.~ Yoshida, Class. Quant. Grav. {\bf 27}, 205006 (2010).
\bibitem {Kraniotis} G.~ V.~ Kraniotis, Class. Quant. Grav. {\bf 28}, 085021 (2011).
\bibitem {JMB} J.~M.~Bardeen, in Black holes, Proceedings of the Les Houches Summer School, Session 215239, edited by C.~ De WWitt and B.~ S.~ De Witt (Gordon and Breach, New York, 1973).
\bibitem {bozzareview} V.~ Bozza, Gen. Relativ. Gravit. {\bf 42}, 2269 (2010).
\bibitem {vih} K. S.~Virbhadra, Phys. Rev. D {\bf 79}, 083004 (2009).
\bibitem {vika} V.~Morozova, B.~Ahmedov, A.~Tursunov, Astrophys. Space Sci. {\bf 346}, 513 (2013).
\bibitem {D} C.~Darwin, Proc. Roy. Soc London A {\bf 249}, 180 (1959).
\bibitem {otros} H.~C.~Ohanian, Am. J. Phys. {\bf 55}, 428 (1987); R.~J.~ Nemiroff, Am. J.
Phys. {\bf 61}, 619 (1993); V. Bozza, S. Capozziello, G. Iovane, and G.
Scarpetta, Gen. Relativ. Gravit. {\bf 33}, 1535 (2001).
\bibitem {claus1} A.~Grenzebach, V.~Perlick, C.~Lämmerzahl, Phys. Rev. D 89, 124004 (2014)
\bibitem {Abdujabbarov} A.~Abdujabbarov, F.~Atamurotov, Y.~Kucukakca, B.~Ahmedov and U.~Camci, Astrophys. Space Sci. {\bf 344}, 429 (2013).
\bibitem {boz} V.~ Bozza, Phys. Rev. D {\bf 66}, 103001 (2002).
\bibitem {Kraniotis14}G.~ V.~ Kraniotis, ArXiv e-prints(2014), 1401.7118.
\bibitem {Falcke} H.~ Falcke, F.~ Melia, and E.~ Agol, Astrophys. J. {\bf 528}, L13 (2000).
\bibitem {bozza2} V.~ Bozza, G.~ Scarpetta, Phys. Rev. D {\bf 76}, 083008 (2007).
\bibitem {chandra} S.~ Chandrasekhar, \textit{The mathematical theory of black holes} (Oxford University Press, New York, 1992).
\bibitem {zakharov05} A.~F.~ Zakharov, A.~A.~ Nucita, F.~ DePaolis, and G.~ Ingrosso, New Astron. {\bf 10}, 479 (2005); A.~F.~ Zakharov, F.~ De Paolis, G.~ Ingrosso, and A.~A.~ Nucita, Astron. Astrophys. {\bf 442}, 795 (2005); F.~ De Paolis, G.~ Ingrosso, A.A. Nucita, A.~ Qadir, and A.~F.~ Zakharov, Gen. Relativ. Gravit. {\bf 43}, 977 (2011).
\bibitem {Ned} P. G.~ Nedkova,  V. K.~ Tinchev, and S. S. Yazadjiev, Phys. Rev. D {\bf 88}, 124019 (2013).
\bibitem {Li1} C.~ Bambi and K.~ Freese, Phys. Rev. D {\bf 79}, 043002 (2009).
\bibitem {amar} L.~ Amarilla, E.~F.~ Eiroa, and G.~ Giribet, Phys. Rev. D {\bf 81}, 124045 (2010).
\bibitem {Hioki08} K.~ Hioki and U.~ Miyamoto, Phys. Rev. D {\bf 78}, 044007 (2008).
\bibitem {Amarilla2} L.~Amarilla and E.~F.~Eiroa, Phys. Rev. D {\bf 85}, 064019 (2012).
\bibitem {schee} J.~ Schee and Z.~ Stuchlik, Int. Jour. Mod. Phys. D {\bf 18}, 983 (2009).
\bibitem {jp} J.~P.~Luminet, Astron.Astrophys. {\bf 75}, 228 (1979).
\bibitem {stu} Z.~ Stuchlik, and J.~ Schee, ArXiv e-print(2014), 1402.2891.
\bibitem {Amarilla1} L.~Amarilla and E.~F.~Eiroa, Phys. Rev. D {\bf 87}, 044057 (2013).
\bibitem {Fa2} F.~Atamurotov, A.~Abdujabbarov and B.~Ahmedov, Astrophys. Space Sci. {\bf 348}, 179 (2013).
\bibitem {Fa3} F.~Atamurotov, A.~Abdujabbarov and B.~Ahmedov, Phys. Rev. D {\bf 88}, 064004 (2013).
\bibitem {Wei} S.~W.~Wei and Y.~X.~Liu, JCAP {\bf 1311}, 063 (2013).
\bibitem {horowitz} G.~T.~Horowitz, \textit{Black Holes in Higher Dimensions}, (Cambridge University Press, 2012).
\bibitem {Emparan} R.~Emparan, H.~S.~Reall, Phys. Rev. Lett. {\bf 88} 101101 (2002); H.~S.~Reall, hep-th/0211290 .
\bibitem {MP} R.~C.~Myers and M.~J.~Perry, Ann. of Phys. {\bf 172} 304 (1986).
\bibitem {Frolov:2003en} V.~P.~Frolov and D.~Stojkovic, Phys. Rev. D {\bf 68}, 064011 (2003).
\bibitem {scht} F.~R.~Tangherlini, Nuovo Cim.\  {\bf 27}, 636 (1963).
\bibitem {Carter68} B.~Carter, Phys. Rev. {\bf 174} 1559 (1968).
\bibitem {VD} V.~Frolov, D.~Stojkovic, Phys. Rev. D {\bf 67}, 024012 (2003).
\bibitem {claus} V.~Diemer, J.~Kunz, C.~Lämmerzahl, S.~Reimers, ArXiv e-prints (2014),1404.3865
\bibitem {Vazquez} S.~ E.~ Vazquez and E.~ P.~ Esteban, Nuovo Cim. {\bf 119}, 489 (2004).
\bibitem {Misner} C.~W.~Misner, K.S.~Thorne, and J.A.~Wheeler, {\em Gravitation} (W.H.~Freeman, San Francisco, 1973).
\bibitem{rp} R. Penrose, {Riv del Nuovo Cimento} {\bf 1}, 252
 (1969); {\it ibid.} in {\it General Relativity}, an Einstein Centenary
 Volume, edited by S. W. Hawking and W. Israel, (Cambridge
 University Press, Cambridge, England, 1979).
\bibitem{r1} P. S. Joshi, {\it Global Aspects in Gravitation and
Cosmology} (Clendron Press, Oxford, 1993).
%
\bibitem {stuchlik14} Z.~ Stuchlik, J.~ Schee, A.~Abdujabbarov,
Phys. Rev. D {\bf 89}, 104048 (2014).
\bibitem {stuchlik13} Z.~ Stuchlik, J.~ Schee, Class. Quant. Grav. {\bf 30}, 075012 (2013).
\bibitem {stuchlik12a} Z.~ Stuchlik, J.~ Schee, Class. Quant. Grav. {\bf 29}, 065002 (2012).
\bibitem {stuchlik12b} Z.~ Stuchlik, J.~ Schee, Class. Quant. Grav. {\bf 27}, 215017 (2010).
%
\bibitem {Patil11} M.~ Patil, P.~ Joshi, Class. Quant. Grav. {\bf 23}, 235012 (2011).
\bibitem {Patil12} M.~ Patil, P.~ Joshi, Phys. Rev. D {\bf 85}, 104014 (2012).
%
\bibitem{Dadhich10}K. Prabhu and N. Dadhich, Phys. Rev. D {\bf 81}, 024011 (2010).
\bibitem{Tsukamoto13}N. Tsukamoto, M. Kimura, T. Harada, Phys. Rev. D {\bf 89},
024020 (2013).
\bibitem {Abdujabbarov13b} A.~Abdujabbarov, N.~Dadhich, B.~Ahmedov, H.~Eshkuvatov, Phys. Rev. D {\bf 88}, 084046 (2013).
%
\bibitem{falcke} H.~ Falcke, S. B. ~Markoff, Class. Quant. Grav. {\bf 30},  244003(2000).
%


\end{references}
\end{document}